# Effect of Spin Polarization on Lattice Vibrations and Electron Wave Interactions in Piezoelectric Semiconductor Quantum Plasma

**Abhishek Yadav and Punit Kumar***

*Department of Physics, University of Lucknow, Lucknow-226007, India*
*Email: kumar_punit@lkouniv.ac.in (Corresponding Author)*

## ABSTRACT

The effect of spin polarization, induced by the difference in concentration of spin-up and spin-down electrons produced under the influence of a magnetic field, on lattice ion vibrations-electron wave interactions, and the resulting amplification of acoustic waves in spin polarised piezoelectric semiconductor quantum plasma has been studied. The dielectric permittivity of the high-density plasma medium has been evaluated through which the dispersion relation has been set up. The gain coefficient of acoustic waves has been obtained using the modified separate spin evolution quantum hydrodynamic (SSE-QHD) model for piezoelectric semiconductor plasma. The study reveals that quantum effects, including Fermi pressure and quantum Bohm potential, reduce wave frequency while spin polarization increases it. Acoustic gain rises significantly with frequency in the presence of quantum effects. Spin polarization also contributes to a slight increase in acoustic wave amplification.

1. **Introduction**

The piezoelectric effect, first identified in solid materials, initiated the scientific investigation of electromechanical coupling at the atomic scale [1]. Subsequent studies reported the alteration of crystal size with an electric field [2-4]. The theory of elasticity, introducing piezoelectric constants, greatly contributed to understanding of the electrical and mechanical aspects of crystals [5]. A linear theory encompassing intrinsic and extrinsic semiconductors, incorporating carrier diffusion, trapping, and drift effects, advanced piezoelectric phenomena in semiconductors was developed [6] leading to extensive industrial applications and energy harvesting [7-9]. The emergence of nanotechnology has expanded the horizons of piezoelectric materials for novel applications [10-12].

Semiconductor plasma systems offer a unique platform where piezoelectric and quantum effects intersect [13,14]. At high densities and low temperatures, electron gas in metals and electron-hole systems in semiconductor plasma display quantum plasma features and shows a transition into the quantum regime. In this unique state, the behaviour of electrons and holes is primarily influenced by quantum effects, resulting in phenomena like quantum degeneracy and quantum confinement. At these scales, quantum effects play a vital role, affecting the transport properties and optical response of semiconductor plasmas. This paves the way for exploring quantum technologies and devices rooted in semiconductor plasma systems [15,16]. Within semiconductors, electrons and holes obey Fermi-Dirac statistics, instead of the classical Boltzmann distribution. For physicists working with quantum structures like quantum wells, wires, and dots, understanding both linear and nonlinear wave characteristics and instabilities driven by carrier dynamics in semiconductors is crucial. In this context, piezoelectric semiconductors stand out as natural choices for converting mechanical stress into an electric field and vice versa. This conversion is enabled by the coupling between lattice ion vibrations and electrokinetic modes through piezoelectricity [17,18].

Piezoelectric semiconductors have emerged as optimal materials for facilitating the interaction between acoustic phonons and plasmons through piezoelectric coupling, making them crucial for various advanced technological applications [19,20]. These semiconductors play a critical role in advancing quantum technologies, particularly in high magnetic field environments. Their integration is crucial for future quantum computing, offering precise qubit control due to their sensitivity to mechanical stress in such environments [21-24]. These materials enhance the performance of power-efficient ultrasound transducers, and improve spectrometer sensitivity in high magnetic field applications [25-29]. In particle detection, their ability to interact with magnetic fields makes them valuable for trapping and manipulating particles in accelerator circuits, facilitating advanced analysis of particle phenomena [30,31].

Phonon-plasmon interactions in semiconductor plasma generate acoustic flux through lattice vibrations [32–35], amplifying acoustic waves when electron drift exceeds sound velocities. This is crucial for applications in acoustic wave amplifiers [36,37], acousto-electric charge transport [38,39], and resonators [40]. Studies have also explored longitudinal acousto-electric wave propagation in colloid-embedded semiconductor plasmas [41], as well as the linear and nonlinear behavior of ion-acoustic waves in three-component quantum plasmas [42] and quantum positron acoustic wave modulation [43].

The rise of quantum technologies has revealed the limitations of classical models for quantum systems, leading to the adoption of quantum models, particularly the quantum hydrodynamic (QHD) model, to explain short-scale collective phenomena in dense plasmas [44,45]. Quantum effects can be investigated through fluid equations, where the Bohm potential captures quantum tunneling and other related phenomena [46-48]. Quantum dielectric tensors and dispersion relations have been derived for both longitudinal and transverse waves in semiconductors [49], with quantum corrections causing rapid decay via Landau damping in longitudinal waves [50]. Quantum surface modes emerge at plasma-vacuum interfaces in magnetized electron-hole semiconductor plasmas [51], and quantum effects lower the threshold electric field for parametric amplification, facilitating the attainment of the required pump electric field, especially in unmagnetized piezoelectric semiconductors [52]. Studies have also examined two-stream instability[53], and reduced Debye length, creating a quasi-quantum lattice of colloid ions in quantum semiconductor plasmas [54]. Additionally, instability in electron beam-pumped GaAs semiconductors can be attributed to the excitation of electron-hole pairs [55]. The QHD model is preferred for its numerical efficiency, direct use of macroscopic variables, and easier boundary condition implementation, making it ideal for studying nonlinear quantum plasma phenomena [56-58].

The intrinsic spin effect of electrons is crucial in quantum plasma, persisting even when macroscopic variations exceed the thermal de Broglie wavelength [59-61]. In high-density plasma, quantum features due to the intrinsic magnetic moment of the electron become noticeable and are found to be somewhat different from those of non-spin quantum effects in plasma [62-63]. Over the past decade, studies have focused on the influence of the spin-1/2 effect on plasma dynamics [64-65]. Previously all the studies were conducted by assuming a macroscopic average of spin 1/2 for electrons. It does not contain any explicit difference between the spin-up and spin-down states of particles, which is a violation of Pauli's exclusion principle. The (SSE-QHD) model addresses the limitations of earlier models by distinguishing between spin-up and spin-down electrons, treating them as independent species and accounting for their interactions [66-67]. High-intensity electromagnetic waves in strongly magnetized quantum plasma induce spin polarization, leading to unequal concentrations of spin-up and spin-down electrons. This model has been used to study even harmonic conversion of whistler pulses [68], SPW excitation in semiconductor plasma [60], and spin-polarization effects on beam-plasma streaming instability [70]. Collective spin effects significantly influence wave propagation in magnetized quantum plasmas, relevant in both laboratory and astrophysical contexts [71-75].

The present work focuses on exploring the coupling between lattice ion vibrations and electron waves in a spin polarized piezoelectric semiconductor quantum medium. The primary objective is to explore the potential amplification of acoustic waves in this environment and to analyze the gain profiles of these waves under varying spin polarization and other quantum effects, which have remained relatively unexplored in previous studies. To achieve this, we have used the SSE-QHD model. Our investigation concentrates on a heavily doped extrinsic n-InSb semiconductor as the medium, where impurities play a more substantial role than carriers from inter-band excitations [76,77]. This study on coupling between lattice ion vibrations and electron waves with spin-polarization (produced under the influence of the external magnetic field) in piezoelectric semiconductor quantum medium has not been reported in the literature so far and the findings are equally applicable to other piezoelectric semiconductors with similar characteristics.

In Section 2, the fundamental theoretical formulation required for the study of the coupled lattice-electron mode in piezoelectric semiconductor quantum plasma has been built up.

This encompasses the relevant equations and models employed to depict the system. The interaction model of e. m. wave with quantum plasma has been developed and the dielectric permittivity of the medium has been obtained by applying the separate spin evolution quantum hydrodynamic (SSE-QHD) model. In Section 3, a quantum dispersion relation has been derived for the coupled lattice-electron mode. In section 4, acoustic wave amplification is thoroughly examined. Finally, Section 5 is devoted to summary and discussion.

## 2. Formalism

Consider a circularly polarized electromagnetic (e. m.) wave propagating parallel to an externally applied static magnetic field $\left(\vec{B} = b\hat{z}\right)$, in the uniform quantum plasma. The fields of the circularly polarized e. m. wave are,

$$\vec{E} = E_0\left(\hat{x} + i\hat{y}\right) e^{i(kz - \omega t)}, \tag{1}$$

$$\vec{B} = \frac{\vec{k} \times \vec{E}}{\omega}. \tag{2}$$

The SSE-QHD fluid equations describing the motion of plasma species under the influence of an electromagnetic field are given as [66-70],

$$\frac{\partial \vec{v}_{j\alpha}}{\partial t} + \vec{v}_{j\alpha} \cdot \nabla \vec{v}_{j\alpha} = -\frac{e_j}{m_j}\left(\vec{E} + \vec{v}_{j\alpha} \times \vec{B}\right) - f\left(\vec{v}_{j\alpha}\right) - \frac{\nabla P_{Fj\alpha}}{m_j n_{j\alpha}} + \frac{\hbar^2}{2m_j^2} \nabla\left(\frac{1}{\sqrt{n_{j\alpha}}} \nabla^2 \sqrt{n_{j\alpha}}\right) + \frac{2\mu}{m_j \hbar} S_{j\alpha}(\nabla B), \tag{3}$$

$$\frac{\partial n_{j\alpha}}{\partial t} + \nabla \cdot \left(n_{j\alpha} \vec{v}_{j\alpha}\right) = 0, \tag{4}$$

$$\frac{d\vec{S}_{j\alpha}}{dt} = -\frac{2\mu}{\hbar}\left(\vec{B} \times \vec{S}_{j\alpha}\right). \tag{5}$$

Eqs. (3), (4) and (5) correspond to momentum, continuity and spin evolution equations for plasma species $j$, $\hbar$ is the reduced Planck's constant, $\mu = -g\mu_B/2$ where $\mu_B = e\hbar/2m_j$ is the Bohr Magneton, $g$ is the Lande's g-factor, $c$ is the speed of light, $f$ is the collision frequency, $\vec{S}_{j\alpha}$ is the spin magnetic moment, $\alpha = \uparrow$ and $\downarrow$ denotes spin up and spin down electrons respectively, $n_{j\alpha}$ is the particle density, $e_j$ is the charge on the particle and $m_j$ is the particle's effective mass. The second term, on the left-hand side of eq. (3) is the convective derivative of the velocity. The first term on RHS of eq. (3) is the Lorentz force under the influence of applied fields, the second term accounts for collisions, the third term is the force due to Fermi pressure $\left( P_{Fj\alpha} = m_e v_{Fj\alpha}^2 n_{j\alpha}^{5/3} / 5 n_{0j\alpha}^{2/3} \right)$, with $v_{Fj\alpha} = \sqrt{\zeta_{3D}}\, \hbar \left( 3\pi^2 n_{0j\alpha} \right)^{1/3} / m_j$ being the Fermi velocity and $\zeta_{3D} = \left[ (1-\eta)^{5/3} + (1+\eta)^{5/3} \right] / 2$, where $\eta = \Delta n_{j\alpha}/n_{0j} = \sum_j (n_\uparrow - n_\downarrow)/n_{0j}$ is the spin polarization due to the magnetic field, with $\Delta n_{j\alpha} = \sum_j (n_\uparrow - n_\downarrow)$ refering to the concentration difference of spin-up and spin-down particles, and the fourth term corresponds to the force of the quantum Bohm potential arising from quantum corrections in density fluctuations and effects the phase and group velocities in semiconductor plasma [78,79]. The last term is the force due to the spin magnetic moment of plasma electrons.

The piezoelectric equations of state are,

$$\vec{T} = \tau \vec{S} - \beta \vec{E}, \tag{6}$$

$$\vec{D} = \varepsilon \vec{E} + \beta \vec{S}. \tag{7}$$

In equations (6) and (7), $\vec{T}, \tau, \vec{S}, \beta, \vec{E}$ and $\vec{D}$ are stress, elastic constant, strain, piezoelectric coupling parameter, electric field and electric displacement respectively. Eq. (6) signifies that an applied electric field causes a strain in the plasma medium, effectively driving mechanical stress which induces a change in electric displacement associated with generation of piezoelectric current [80-82].

The strain $\vec{S} = \partial \vec{u}/\partial x$, with $u$ being the physical displacement of particles. Equation (6), leads to the wave equation in an elastic medium,

$$\frac{\partial^2 \vec{u}}{\partial t^2} = c_s^2 \frac{\partial^2 \vec{u}}{\partial x^2} - \frac{\beta}{\rho} \nabla \cdot \vec{E} - \frac{\beta}{\rho} \nabla \cdot \left( \vec{v}_{j\alpha} \times \vec{B} \right). \tag{8}$$

where, $c_s = \sqrt{\tau/\rho}$ is the speed of sound and $\rho$ is the density of the elastic medium.

The perturbative expansion of eqs. (3), (4) and (5) in the first order of the e. m. field gives,

$$\frac{\partial \vec{v}_{e\alpha}^{(1)}}{\partial t} = -\frac{e}{m_e}\vec{E}^{(1)} - \frac{e}{m_e}\left(\vec{v}_{e\alpha}^{(1)} \times \vec{B}_0 - \vec{v}_{0e\alpha} \times \vec{B}^{(1)}\right) - f\left(\vec{v}_{e\alpha}^{(1)}\right) - \frac{\nabla P_{Fe\alpha}^{(1)}}{m_e n_{0e\alpha}} + \frac{h^2}{4m_e^2 n_{0e\alpha}}\nabla\left(\nabla^2 n_{e\alpha}^{(1)}\right)$$
$$+ \frac{2\mu}{m_e h}S_{0\alpha}\left(\nabla B^{(1)}\right),$$

(9)

$$\frac{\partial n_{e\alpha}^{(1)}}{\partial t} + n_{0e\alpha}\nabla\cdot\left(\vec{v}_{e\alpha}^{(1)}\right) = 0, \tag{10}$$

$$\frac{\partial \vec{S}_\alpha^{(1)}}{\partial t} = -\frac{2\mu}{h}\left(\vec{B}^{(1)} \times \vec{S}_{0\alpha} + \vec{B}_0 \times \vec{S}_\alpha^{(1)}\right). \tag{11}$$

Assuming all the perturbed quantities to vary as $e^{i(kz-\omega t)}$, we get the first order perturbed eqs. for transverse quiver velocity components,

$$v_{e\alpha x}^{(1)} = \frac{-i\frac{e}{m_e}\left(\frac{k^2 Q_{e\alpha}}{\omega} - \omega - if\right)E_x^{(1)} + \left(\frac{e}{m_e}\right)^2 bE_y^{(1)} - \frac{2\mu e b}{m_e^2 h}S_{0\alpha y}\left(\nabla B_y^{(1)}\right) + i\left(\frac{k^2 Q_{e\alpha}}{\omega} - \omega - if\right)\frac{2\mu}{m_e h}S_{0\alpha x}\left(\nabla B_x^{(1)}\right)}{\left[\omega_{ce}^2 - \left(\frac{k^2 Q_{e\alpha}}{\omega} - \omega - if\right)^2\right]},$$

(12)

and

$$v_{e\alpha y}^{(1)} = \frac{-i\frac{e}{m_e}\left(\frac{k^2 Q_{e\alpha}}{\omega} - \omega - if\right)E_y^{(1)} - \left(\frac{e}{m_e}\right)^2 bE_x^{(1)} + \frac{2\mu e b}{m_e^2 h}S_{0\alpha x}\left(\nabla B_x^{(1)}\right) + i\left(\frac{k^2 Q_{e\alpha}}{\omega} - \omega - if\right)\frac{2\mu}{m_e h}S_{0\alpha y}\left(\nabla B_y^{(1)}\right)}{\left[\omega_{ce}^2 - \left(\frac{k^2 Q_{e\alpha}}{\omega} - \omega - if\right)^2\right]},$$

(13)

where, $Q_{e\alpha} = v_{F\alpha}^2\left(1 + \frac{k^2 v_{F\alpha}^2}{4\omega_{p\alpha}^2}H^2\right).$

For plasma electrons, the current density which is the sum of conduction and magnetization current densities is given as,

$$\vec{J}_{e\alpha} = \vec{J}_{cj\alpha} + \vec{J}_{Mj\alpha} = \hat{\sigma}.\vec{E} \tag{14}$$

where, $\vec{J}_{cj\alpha}\left(=\sum_{j\alpha} en_{0j\alpha} v_{j\alpha}\right)$ and $\vec{J}_{Mj\alpha}\left(=\sum_{j\alpha} \nabla \times M_{j\alpha}\right)$ are the conventional and magnetization current densities respectively. Magnetization current density arises from the motion of charged particles around magnetic field lines. This circulation of current is linked to the quantized energy levels, which are influenced by the particle's spin arising from the quantization of spin thereby giving rise to unique current patterns. $M_{j\alpha} = \dfrac{2n_{j\alpha} \mu_B \vec{S}_\alpha}{\hbar}$ being the magnetization due to spin, and $\hat{\sigma}$ is the conductivity of the plasma medium.

The conduction current density of plasma electrons,

$$\vec{J}_{ce\alpha}^{(1)} = -en_{0e\alpha} v_{e\alpha}^{(1)} \tag{15}$$

whose components are,

$$J_{ce\alpha x}^{(1)} = -en_{0\alpha} \left[ \dfrac{-i\dfrac{e}{m_e}\left(\dfrac{k^2 Q_{e\alpha}}{\omega} - \omega - if\right) E_x^{(1)} + \left(\dfrac{e}{m_e}\right)^2 \left[ bE_y^{(1)} - \dfrac{2\mu eb}{m_e^2 \hbar} S_{0\alpha y}(\nabla B_y^{(1)}) \right] + i\left(\dfrac{k^2 Q_{e\alpha}}{\omega} - \omega - if\right)\dfrac{2\mu}{m_e \hbar} S_{0\alpha x}(\nabla B_x^{(1)})}{\left[\omega_{ce}^2 - \left(\dfrac{k^2 Q_{e\alpha}}{\omega} - \omega - if\right)^2\right]} \right],$$

$$\tag{16}$$

$$J_{ce\alpha y}^{(1)} = -en_0 \left[ \dfrac{-i\dfrac{e}{m_e}\left(\dfrac{k^2 Q_{e\alpha}}{\omega} - \omega - if\right) E_y^{(1)} - \left(\dfrac{e}{m_e}\right)^2 \left[ bE_x^{(1)} + \dfrac{2\mu eb}{m_e^2 \hbar} S_{0\alpha x}(\nabla B_x^{(1)}) \right] + i\left(\dfrac{k^2 Q_{e\alpha}}{\omega} - \omega - if\right)\dfrac{2\mu}{m_e \hbar} S_{0\alpha y}(\nabla B_y^{(1)})}{\left[\omega_{ce}^2 - \left(\dfrac{k^2 Q_{e\alpha}}{\omega} - \omega - if\right)^2\right]} \right],$$

$$\tag{17}$$

and

$$J^{(1)}_{ce\alpha z} = 0. \tag{18}$$

The first order spin magnetic moment for plasma electrons can be obtained by solving equation (11) as,

$$S^{(1)}_{\alpha x} = \frac{\frac{2\mu}{h} S_{0_{\alpha z}} \left( i\omega B^{(1)}_y - \frac{2\mu}{h} B_{0_z} B^{(1)}_x \right)}{\left[ \omega^2 - \left( \frac{2\mu}{h} B_{0_z} \right)^2 \right]}, \tag{19}$$

$$S^{(1)}_{\alpha y} = \frac{\frac{2\mu}{h} S_{0_{\alpha z}} \left( -i\omega B^{(1)}_x - \frac{2\mu}{h} B_{0_z} B^{(1)}_y \right)}{\left[ \omega^2 - \left( \frac{2\mu}{h} B_{0_z} \right)^2 \right]}, \tag{20}$$

and

$$S^{(1)}_{\alpha z} = -\frac{2\mu}{i\omega h} \left( B^{(1)}_x S_{0_{\alpha y}} - B^{(1)}_y S_{0_{\alpha x}} \right). \tag{21}$$

Considering $\vec{S}_\uparrow = -\vec{S}_\downarrow$ [83,84], the magnetization current density is given as

$$\vec{J}_{M\alpha} = \nabla \times \left( \vec{M}_\uparrow + \vec{M}_\downarrow \right) = \frac{2\mu}{\hbar} \nabla \times \left( n_\uparrow \vec{S}_\uparrow + n_\downarrow \vec{S}_\downarrow \right) = \frac{2\mu}{\hbar} \left( n_\uparrow - n_\downarrow \right) \nabla \times \vec{S}_\uparrow. \tag{22}$$

In the non-degenerate plasma $(T \gg T_F)$, $n_\uparrow - n_\downarrow = n \tanh(\mu B/k_B T_F)$, and in degenerate plasma $(T \gg T_F)$, $n_\uparrow - n_\downarrow = (3n \mu B_0 / 2k_B T_F)$, where $T_F$ is the Fermi temperature of the electrons. The x, y and z components of equation (22) are

$$J^{(1)}_{M\alpha x} = -\left( \frac{2\mu}{\hbar} \right)^2 \frac{\left( n_\uparrow - n_\downarrow \right) S_{0_{\alpha z}}}{\left[ \omega^2 - \left( \frac{2\mu}{\hbar} B_{0_z} \right)^2 \right]} \nabla \times \left( i\omega B^{(1)}_y - \frac{2\mu}{\hbar} B_{0_z} B^{(1)}_x \right), \tag{23}$$

$$J^{(1)}_{M\alpha y} = -\left(\frac{2\mu}{\hbar}\right)^2 \frac{(n_\uparrow - n_\downarrow)S_{0\alpha z}}{\left[\omega^2 - \left(\frac{2\mu}{\hbar}B_{0_z}\right)^2\right]} \nabla \times \left(-i\omega B^{(1)}_x - \frac{2\mu}{\hbar}B_{0_z}B^{(1)}_y\right), \tag{24}$$

and

$$J^{(1)}_{M\alpha z} = \left(\frac{2\mu}{\hbar}\right)^2 \frac{(n_\uparrow - n_\downarrow)}{i\omega} \nabla \times \left(B^{(1)}_x S_{0\alpha y} - B^{(1)}_y S_{0\alpha x}\right). \tag{25}$$

In a similar manner, the first order dynamical eqs. for ion are,

$$\frac{\partial \vec{v}^{(1)}_i}{\partial t} = \frac{e}{m_i}\vec{E}^{(1)} + \frac{e}{m_i}\left(\vec{v}^{(1)}_i \times \vec{B}_0 - \vec{v}_{0i} \times \vec{B}^{(1)}\right), \tag{26}$$

$$\frac{\partial n^{(1)}_i}{\partial t} + n_{0i}\nabla \cdot \left(\vec{v}^{(1)}_i\right) = 0, \tag{27}$$

and

$$\frac{\partial^2 \vec{u}^{(1)}_i}{\partial t^2} = c_s^2 \frac{\partial^2 \vec{u}^{(1)}_i}{\partial x^2} - \frac{\beta}{\rho_i}\nabla \cdot \vec{E}^{(1)} - \frac{\beta}{\rho_i}\nabla \cdot \left(\vec{v}^{(1)}_i \times \vec{B}_0 - \vec{v}_{0i} \times \vec{B}^{(1)}\right). \tag{28}$$

The neglect of quantum terms such as force due to Fermi pressure, Bohm potential and spin magnetic moment plasma electrons in ion dynamics within the context of quantum plasma is based on considerations that ions are much heavier than electrons i.e., $m_e/m_i = 1/1836$. Ions, being much heavier, are treated classically while electrons quantum mechanically. The thermal de Broglie wavelength of ions doesn't become comparable to the plasma Debye length, justifying their classical treatment in the equations of motion [85,86]. From the above equations, equations of motion for lattice ions come out to be,

$$\left(\omega^2 - c_s^2 k^2\right)u_{ix} = \frac{\beta}{\rho_i}ik\left(\frac{(\omega + if)^2}{\left((\omega + if)^2 - \omega_{ci}^2\right)}E^{(1)}_x + \frac{i(\omega + if)\omega_{ci}}{\left((\omega + if)^2 - \omega_{ci}^2\right)}E^{(1)}_y\right), \tag{29}$$

$$\left(\omega^2 - c_s^2 k^2\right) u_{iy} = \frac{\beta}{\rho_i} ik \left(\frac{(\omega+if)^2}{\left((\omega+if)^2 - \omega_{ci}^2\right)} E_y^{(1)} - \frac{i(\omega+if)\omega_{ci}}{\left((\omega+if)^2 - \omega_{ci}^2\right)} E_x^{(1)}\right), \tag{30}$$

and

$$\left(\omega^2 - c_s^2 k^2\right) u_{iz} = 0. \tag{31}$$

The above eqs. describe the motion of lattice ions under the influence of e. m. field, incorporating piezoelectric coupling. The piezoelectric coupling constant $\beta$ links the electric field to ion displacement.

### 3. Coupling of modes

We now proceed to study the coupling of electron-ion modes due to the piezoelectric field in n-type piezoelectric semiconductor quantum plasma. The dielectric permittivity $\hat{\varepsilon}_r = \hat{I} + i\sigma/\varepsilon_0 \omega$ of the medium is obtained using eq. (14),

$$\hat{\varepsilon}_r = \begin{pmatrix} \varepsilon_{11} & \varepsilon_{12} & \varepsilon_{13} \\ \varepsilon_{21} & \varepsilon_{22} & \varepsilon_{23} \\ \varepsilon_{31} & \varepsilon_{32} & \varepsilon_{33} \end{pmatrix} \tag{32}$$

where,

$$\varepsilon_{11} = \varepsilon_{22} = \varepsilon_L - \frac{(\beta k)^2}{\varepsilon_0 \rho_i (\omega^2 - c_s^2 k^2)} \cdot \frac{(\omega+if)^2}{\left((\omega+if)^2 - \omega_{ci}^2\right)} - \frac{1}{\omega \varepsilon_0} \left(\frac{\omega_{p\alpha}^2 \varepsilon \left(\frac{k^2 Q_{e\alpha}}{\omega} - \omega - if\right)}{\omega_{ce}^2 - \left(\frac{k^2 Q_{e\alpha}}{\omega} - \omega - if\right)^2}\right)$$

$$+ \frac{i}{\omega \varepsilon_0} \left(\left(\frac{2\mu}{\hbar}\right)^2 \frac{(n_\uparrow - n_\downarrow) S_{0\alpha}}{\left[\omega^2 - \left(\frac{2\mu}{\hbar} b\right)^2\right]} \cdot \left(\frac{2\mu k^2}{\hbar \omega} b\right)\right),$$

$$\varepsilon_{12} = -\varepsilon_{21} = -\frac{(\beta k)^2}{\varepsilon_0 \rho_i (\omega^2 - c_s^2 k^2)} \cdot \frac{i(\omega + if)\omega_{ci}}{((\omega + if)^2 - \omega_{ci}^2)} - \frac{i}{\omega \varepsilon_0}\left(\frac{\omega_{p\alpha}^2 \omega_{ce}\varepsilon}{\left(\omega_{ce}^2 - \left(\frac{k^2 Q_{e\alpha}}{\omega} - \omega - if\right)^2\right)}\right)$$

$$+ \frac{1}{\omega \varepsilon_0}\left(\left(\frac{2\mu k}{\hbar}\right)^2 \frac{(n_\uparrow - n_\downarrow)S_{0_\alpha}}{\left[\omega^2 - \left(\frac{2\mu}{\hbar}b\right)^2\right]}\right),$$

$\varepsilon_{13} = \varepsilon_{31} = \varepsilon_{23} = \varepsilon_{32} = 0$

and

$\varepsilon_{33} = \varepsilon_L.$

The dielectric permittivity of the n-type piezoelectric semiconductor quantum plasma is a critical factor that determines how the medium responds to electromagnetic fields. As shown in the derived expression, the components $\varepsilon_{11}, \varepsilon_{22}, \varepsilon_{12}, \varepsilon_{21}$ are influenced by both quantum mechanical terms and spin effects. The term $Q_{e\alpha}$ highlights the influence of quantum corrections in the electron response, directly modifying the dielectric behavior and the term $(n_\uparrow - n_\downarrow)S_{0_\alpha}$ in the permittivity expression reflects the difference in populations of spin-up and spin-down electrons. This difference contributes to the anisotropy in the dielectric response, particularly in the presence of magnetic fields.

Now by substituting the dielectric permittivity from eq. (32) in wave eq. $\vec{\nabla}^2 \vec{E} - \vec{\nabla}(\vec{\nabla} \cdot \vec{E}) + \frac{\omega^2}{c^2}\varepsilon_r \cdot \vec{E} = 0,$ we obtain the coupled dispersion relation in terms of the quantum parameter $H\left(= \hbar \omega_p / 2 k_B T_F\right),$

$$\left(\omega^2 - c_s^2 k^2\right)\left(\varepsilon_L - \left(\frac{kc}{\omega}\right)^2 - \sum_\alpha \left(\frac{\omega_{p\alpha}^2 \varepsilon_L}{\left(\omega^2 - k^2 v_{F\alpha}^2 \left(1 + \frac{k^2 v_{F\alpha}^2}{4\omega_{p\alpha}^2} H^2\right) + if\omega - \omega\omega_{ce}\right)} - \frac{i}{\omega\varepsilon_0}\left(\frac{2\mu}{\hbar}\right)^2 \frac{(n_\uparrow - n_\downarrow)k^2 S_{0\alpha}}{\omega^2\left(1 + \frac{2\mu b}{\hbar\omega}\right)}\right)\right)$$

$$= \frac{\beta^2 k^2}{\rho_i \varepsilon_0}\left(\frac{(\omega + if)}{(\omega + if) + \omega_{ci}}\right).$$

(33)

In the above equation, the first term on the left-hand side represents the lattice acoustic mode, while the second term represents the electron plasma mode in piezoelectric semiconductor quantum plasma. The term on the right-hand side is the coupling term that accounts for the interaction between the electron plasma and lattice acoustic modes. In the absence of piezoelectricity $(\beta = 0)$, the coupling parameter vanishes and the dispersion relation decouples into two independent modes, the lattice acoustic and the Langmuir modes. The electron plasma and lattice vibrations evolve independently of each other, without any interaction or coupling between them,

$$\left(\varepsilon_L - \left(\frac{kc}{\omega}\right)^2 - \sum_\alpha \left(\frac{\omega_{p\alpha}^2 \varepsilon_L}{\left(\omega^2 - k^2 v_{F\alpha}^2 \left(1 + \frac{k^2 v_{F\alpha}^2}{4\omega_{p\alpha}^2} H^2\right) + if\omega - \omega\omega_{ce}\right)} - \frac{i}{\omega\varepsilon_0}\left(\frac{2\mu}{\hbar}\right)^2 \frac{(n_\uparrow - n_\downarrow)k^2 S_{0\alpha}}{\omega^2\left(1 + \frac{2\mu b}{\hbar\omega}\right)}\right)\right) = 0$$

(34)

and

$$\left(\omega^2 - c_s^2 k^2\right) = 0 \tag{35}$$

In the numerical analysis to follow, the parameter chosen are for an n-type InSb semiconductor; $\varepsilon_0 = 8.85 \times 10^{-12}\, C^2/N.m^2$, $\varepsilon_L = 17.54$, $\varepsilon = 1.55 \times 10^{-10}$, $n_{e0} = 10^{26}/m^3$, $\rho_i = 5.8 \times 10^3$, $m_e = 0.014 m_0$, $m_0 = 9.1 \times 10^{-31}\, Kg$, $c_s = 2500\, m/s$, $T = 77K$ [46-50] and the piezoelectric coupling constant $\beta$ for such type of materials ranges from $0.045\,C/m^2$ to $0.35\,C/m^2$ [52,87-92].

Figure 1 shows the variation of normalised wave frequency $\omega/\omega_p$ with respect to normalized propagation vector $kc/\omega_p$. The solid line shows the variation in quantum plasma, while the dashed line shows the trend in absence of quantum effects $(\hbar \to 0)$. It is evident from the figure that the wave frequency is reduced by about 12.5% in quantum plasma as compared to the case where quantum effects are absent. This is due to the dominance of Fermi pressure over thermal pressure resulting in a higher number of energy levels, which introduces degeneracy, thereby reducing the wave frequency.

Figure 2 shows the variation of normalised wave frequency $\omega/\omega_p$ with normalized propagation vector $kc/\omega_p$ for different values of spin polarization $\eta$. The solid, dashed and dotted line show the variation for $\eta = 1$, $\eta = 0.5$ and $\eta = 0$ respectively. The wave frequency for fully spin-polarized case is about 32.5% more than that of unpolarized plasma at $k \approx 5$. This increase is due to the high value of Fermi pressure caused by spin polarization and the electron's spin magnetic moment.

In figure 3 the variation of normalised wave frequency $\omega/\omega_p$ with normalized propagation vector $kc/\omega_p$ is shown for different values of quantum parameter $H$. The wave frequency increases by 25% for $H = 0.045$ in comparison to $H = 0.030$ and by 14.2% for $H = 0.066$ in comparison to $H = 0.045$. This is due to the concurrent influence of Fermi pressure and quantum Bohm potential, as quantum parameter $H$ shows their combined effect. Thus, we conclude that the transmission of power increases for the same wave transmission due to quantum effects involving quantum tunneling.

4. **Acoustic gain**

Assuming the quantum plasma in collision dominated regime, under standard approximation $(kc_s/\omega) = (1 + i\alpha)$ [33], the dispersion relation in eq. (33) can be solved to obtain acoustic gain $\Gamma$ as,

$$\Gamma = \sum_{\alpha} \frac{\left(\dfrac{\omega\varepsilon_L \beta^2 \omega_{p\alpha}^2}{2c\varepsilon_0 f}\right)}{\left[\dfrac{\omega_{p\alpha}^2 \varepsilon_L}{f} - \left(k^2 D_\alpha + \dfrac{\omega\omega_{ce}}{f}\right)\left(\varepsilon_L - \dfrac{k^2 c^2}{\omega^2}\right) - \dfrac{1}{\varepsilon_0}\left(\dfrac{2\mu}{\hbar}\right)^2 \dfrac{(n_\uparrow - n_\downarrow)k^2 S_{0\alpha}}{\omega^2 \left(1 + \dfrac{2\mu b}{\hbar\omega}\right)}\right]^2}$$

$$+ \left[\omega\left(\varepsilon_L - \dfrac{k^2 c^2}{\omega^2} + \dfrac{1}{\omega^2 \varepsilon_0}\left(k^2 D_\alpha + \dfrac{\omega\omega_{ce}}{f}\right)\left(\dfrac{2\mu}{\hbar}\right)^2 \dfrac{(n_\uparrow - n_\downarrow)k^2 S_{0\alpha}}{\omega^2 \left(1 + \dfrac{2\mu b}{\hbar\omega}\right)}\right)\right]^2 \quad (36)$$

where, $D_\alpha = f \cdot D_{f\alpha}\left(1 + \dfrac{k^2 v_{F\alpha}^2}{4\omega_{p\alpha}^2} H^2\right)$, and $D_{f\alpha} = \dfrac{v_{F\alpha}^2}{f}$ are diffusion constants.

Figure 4 shows the variation of the gain profile of acoustic wave with frequency in presence and absence of quantum effects $(\hbar \to 0)$. Both the curves show an increase in gain with frequency, but the quantum gain rises much more sharply. This difference highlights the impact of quantum effects, such as Fermi pressure and Bohm potential, enhance the coherence and energy transfer mechanisms in quantum plasma, resulting in a higher gain compared to the case where quantum effects are absent.

Figure 5 shows the changes occurred in gain profiles of acoustic wave by varying frequency for different values of spin polarization $\eta$. It is observed from this figure that there is a slight increase in the magnitude of acoustic gain as the spin polarization decreases. At higher frequencies, the effect of spin polarization is more pronounced, as shown by the growing separation between the curves. This suggests that spin effects are particularly important in high-frequency domains.

Figure 6 shows the plot of the gain profile versus piezoelectric coupling strength for different values of quantum parameter $H$. It can be seen from figure that the gain grows more steeply as piezoelectric coupling strength increases and attains higher value for larger value of quantum parameter, suggesting a strong coupling between the piezoelectric effect and quantum properties at higher coupling strengths.

## 5. Summary and discussion

This study explores the influence of spin polarization, induced by the concentration difference of spin-up and spin-down electrons under a magnetic field, on the interactions between lattice ion vibrations and electron waves, focusing on the amplification of acoustic waves in spin-polarized piezoelectric semiconductor quantum plasma. By evaluating the dielectric permittivity of this high-density plasma medium, we establish a quantum-modified dispersion relation and derive the expression of gain for acoustic waves using the modified separate spin evolution quantum hydrodynamic (SSE-QHD) model.

Our analysis reveals that the normalized wave frequency decreases in quantum plasma due to the dominance of Fermi pressure, which introduces degeneracy. Higher spin polarization increases the wave frequency, with fully spin-polarized plasma showing approximately 32.5% increase. The wave frequency also rises with the quantum parameter, demonstrating concurrent influences of Fermi pressure and quantum Bohm potential, showing up to a 25% increase. Additionally, acoustic wave gain increases more rapidly with frequency in the presence of quantum effects. Spin polarization slightly increases the magnitude of acoustic gain, while higher piezoelectric coupling strength and quantum parameter values lead to greater acoustic gain. The findings indicate that quantum effects, particularly Fermi pressure and quantum Bohm potential, significantly affect wave frequency and acoustic gain in spin-polarized piezoelectric semiconductor quantum plasma. Spin polarization further enhances these properties by increasing Fermi pressure and utilizing the electron's spin magnetic moment. These quantum phenomena collectively enhance power transmission for the same wave, showing the importance of quantum effects in high-density plasma media.

The findings of this paper have several promising applications in cutting-edge technologies. In quantum computing and spintronics, the amplified acoustic waves and spin polarization effects could be utilized to improve qubit manipulation, enabling more precise control in quantum processors operating under high magnetic fields. This enhanced acoustic wave gain can improve high-frequency ultrasound imaging by boosting the sensitivity and resolution of ultrasound transducers. The ability to amplify acoustic waves in spin-polarized quantum plasma also has the potential to improve particle detection systems in accelerator circuits, providing more accurate measurements and aiding in the analysis of particle interactions and quantum processes.

**Declaration of Competing Interest**

The authors report no declarations of interest.

**Acknowledgement**

The authors thank SERB – DST, Govt. of India for financial support under MATRICS scheme (grant no. : MTR/2021/000471).

**References**

[1] Curie J, Curie P. Développement, par pression, de l'électricité polaire dans les cristaux hémièdres à faces inclines. Comptes Rendus 1880;91:294.

[2] Lippmann G. Relations entre les phénomènes électriques et capillaires. Ann Chim Phys 1875;5:494.

[3] Cady WG. Piezoelectricity: An introduction to the theory and applications of electromechanical phenomena in crystals. McGraw-Hill; 1946.

[4] Mason WP. Piezoelectricity, its history and applications. J Acoust Soc Am 1981;70:1561.

[5] Voigt W. On an apparently necessary extension of the theory of elasticity. Ann Phys 1894;52:536.

[6] Hutson AR. Piezoelectricity and conductivity in ZnO and CdS. Phys Rev Lett 1960;4:505.

[7] Roundy S, Wright PK. A piezoelectric vibration-based generator for wireless electronics. Smart Mater Struct 2004;13:1131.

[8] Lefeuvre E, Badel A, Richard C, Guyomar D. Piezoelectric energy harvesting device optimization by synchronous electric charge extraction. J Intell Mater Syst Struct 2005;16:865.

[9] Sodano HA, Inman DJ, Park G. Comparison of piezoelectric energy harvesting devices for recharging batteries. J Intell Mater Syst Struct 2005;16:799–807.

[10] Cook-Chennault KA, Thambi N, Sastry AM. Powering MEMS portable devices—a review of non-regenerative and regenerative power supply systems with special emphasis on piezoelectric energy harvesting systems. Smart Mater Struct 2008;17:043001.

[11] Wang ZL, Wang X, Song J. Piezoelectric nanogenerators for self-powered nanodevices. IEEE Pervasive Comput 2008;7:49–55.

[12] Li P, Jin F, Ma J. One-dimensional dynamic equations of a piezoelectric semiconductor beam with a rectangular cross-section and their application in static and dynamic characteristic analysis. Appl Math Mech 2018;39:685.

[13] Schäfer W, Wegener M. Semiconductor Optics and Transport Phenomena. Springer; 2002.

[14] Haug H, Jauho AP. Quantum Kinetics in Transport and Optics of Semiconductors. Springer; 2008.

[15] Haug H, Koch SW. Quantum Theory of the Optical and Electronic Properties of Semiconductors. World Scientific; 2009.

[16] Bonitz M, Horing N, Ludwig P. Introduction to Complex Plasmas. Springer; 2010.

[17] Anton SR, Sodano HA. A review of power harvesting using piezoelectric materials (2003–2006). Smart Mater Struct 2007;16:R1.

[18] Twiefel J, Westermann H. Survey on broadband techniques for vibration energy harvesting. J Intell Mater Syst Struct 2013;24:1291–1302.

[19] Sodano HA, Inman DJ. A review of power harvesting from vibration using piezoelectric materials. Smart Mater Struct 2007;16:R1–21.

[20] Twiefel J, Westermann H. Survey on broadband techniques for vibration energy harvesting. J Intell Mater Syst Struct 2013;24:1291–1302.

[21] Seo H, Govoni M, Galli G. Design of defect spins in piezoelectric aluminum nitride for solid-state hybrid quantum technologies. Sci Rep 2016;6:20803.

[22] Teo KH, Zhang Y, Chowdhury N, Rakheja S, Ma R, Xie Q, Palacios T. Emerging GaN technologies for power, RF, digital, and quantum computing applications: Recent advances and prospects. J Appl Phys 2021;130.

[23] Salfi J, Tong M, Rogge S, Culcer D. Quantum computing with acceptor spins in silicon. Nanotechnology 2016;27:244001.

[24] Ladd T, Jelezko F, Laflamme R, et al. Quantum computers. Nature 2010;464:45–53.

[25] Rugar D, Budakian R, Mamin H, et al. Single spin detection by magnetic resonance force microscopy. Nature 2004;430:329–332.

[26] Lethiecq M, Berson M, Feuillard G, Patat F. Principles and applications of high-frequency medical imaging. Adv Acoust Microsc 1996:39–102.

[27] Harvey G, Gachagan A, Mutasa T. Review of high-power ultrasound-industrial applications and measurement methods. IEEE Trans Ultrason Ferroelectr Freq Control 2014;61:481–495.

[28] Wang DJ, Leigh JS. Wireless precision piezoelectric thermometer using an RF excitation-detection technique with an NMR probe. J Magn Reson B 1994;105:25–30.

[29] Smoleński T, Dolgirev PE, Kuhlenkamp C, et al. Signatures of Wigner crystal of electrons in a monolayer semiconductor. Nature 2021;595:53–57.

[30] Marechal P, Levassort F, Holc J, Tran-Huu-Hue LP, Kosec M, Lethiecq M. High-frequency transducers based on integrated piezoelectric thick films for medical imaging. IEEE Trans Ultrason Ferroelectr Freq Control 2006;53:1524.

[31] Wendt O, Oellinger J, Lüth TC, Felix R, Boenick U. The effects of the use of piezoelectric motors in a 1.5-Tesla high-field magnetic resonance imaging system. Biomed Tech 2000;45:20.

[32] Mosekilde E. Quantum theory of acoustoelectric interaction. Phys Rev B 1974;9:682–689.

[33] Dey M, Ghosh S. Amplification of acoustic waves in magnetized high resistivity piezoelectric semiconductors. Phys Status Solidi B 1990;157:159–166.

[34] Alok U, Kumar M, Singh BK, Kumar P. Acoustoelectric effect in semiconductor superlattice. Bibechana 2012;8:67–72.

[35] Dubey P, Ghosh S. Modified longitudinal phonon-plasmon interactions in nanoparticle-doped piezoelectric semiconductors. Acta Acust United AC 2016;102:436–440.

[36] White DL. Amplification of ultrasonic waves in piezoelectric semiconductors. J Appl Phys 1962;33:2547–2554.

[37] Yang JS, Zhou HG. Acoustoelectric amplification of piezoelectric surface waves. Acta Mech 2004;172:113–122.

[38] Schulein FJR, Muller K, Bichler M, et al. Acoustically regulated carrier injection into a single optically active quantum dot. Phys Rev B 2013;88:085307.

[39] Buyukkose S, Hernandez-Minguez A, Vratzov B, et al. High-frequency acoustic charge transport in GaAs nanowires. Nanotechnology 2014;25:135204.

[40] Ghokhale VJ, Zadeh MR. Phonon-electron interactions in piezoelectric semiconductor bulk acoustic wave resonator. Sci Rep 2014;4.

[41] Sharma A, Yadav N, Ghosh S. Modified acousto-electric interactions in colloids laden semiconductor quantum plasmas. Int J Sci Res Publ 2013;3:1.

[42] Ali S, Moslem WM, Shukla PK, Schlickeiser R. Linear and nonlinear ion-acoustic waves in an unmagnetized electron-positron-ion quantum plasma. Phys Plasmas 2007;14:082307.

[43] Amin MR. Modulation of a quantum positron acoustic wave. Astrophys Space Sci 2015;359:1.

[44] Bret A. Filamentation instability in a quantum plasma. Phys Plasmas 2007;14:084503.

[45] Haas F, Bret A. Nonlinear low-frequency collisional quantum Buneman instability. Europhys Lett 2012;97:26001.

[46] Manfredi G, Hervieux PA, Hurst J. Fluid descriptions of quantum plasmas. Rev Mod Plasma Phys 2021;5:1.

[47] Manfredi G, Haas F. Self-consistent fluid model for a quantum electron gas. Phys Rev B 2001;64:075316.

[48] Haas F, Garcia LG, Goeder J, Manfredi G. Quantum ion-acoustic waves. Phys Plasmas 2003;10:3858–3866.

[49] Kumar P, Singh S, Ahmad N. Beam-plasma streaming instability in spin polarized quantum magnetoplasma. Phys Scr 2020;95:075604.

[50] Mehramiz A, Mahmoodi J, Sobhanian S. Approximation method for a spherical bound system in the quantum plasma. Phys Plasmas 2010;17:082110.

[51] Misra AP. Electromagnetic surface modes in a magnetized quantum electron-hole plasma. Phys Rev E 2011;83:057401.

[52] Ghosh S, Dubey S, Vanshpal R. Quantum effect on parametric amplification characteristics in piezoelectric semiconductors. Phys Lett A 2010;375:43.

[53] Zeba I, Yahi ME, Shukla PK, Moslem WM. Electron–hole two-stream instability in a quantum semiconductor plasma with exchange-correlation effects. Phys Lett A 2012;376:2309.

[54] Zeba I, Uzma C, Jamil M, Salimullah M. Colloidal crystal formation in a semiconductor quantum plasma. Phys Plasmas 2010;17:032105.

[55] Yahia ME, Azzouz IM, Moslem WM. Quantum effects in electron beam pumped GaAs. Appl Phys Lett 2013;103:082105.

[56] Vladimirov SV, Tyshetskiy YO. On description of a collisionless quantum plasma. Phys Usp 2011;54:1243.

[57] Shukla PK, Eliasson B. Nonlinear aspects of quantum plasma physics. Phys Usp 2010;53:51.

[58] Shukla PK, Ali S, Stenflo L, Marklund M. Nonlinear wave interactions in quantum magnetoplasmas. Phys Plasmas 2006;13:112111.

[59] Polyakov PA. Hydrodynamic description of plasma waves including electron spin. Sov Phys J 1979;22:310–312.

[60] Marklund M, Brodin G. Dynamics of spin-1/2 quantum plasmas. Phys Rev Lett 2007;98:025001.

[61] Brodin G, Marklund M. On the possibility of metamaterial properties in spin plasmas. New J Phys 2008;10:115031.

[62] Rastunkov VS, Krainov VP. Relativistic electron drift in overdense plasma produced by a superintense femtosecond laser pulse. Phys Rev E 2004;69:037402.

[63] Manfredi G. How to model quantum plasmas. Fields Inst Commun 2005;46:263–287.

[64] Shahid M, Melrose DB, Jamil M, Murtaza G. Spin effect on parametric interactions of waves in magnetoplasmas. Phys Plasmas 2012;19:112114.

[65] Hu QL, Linzhou S, Yu X, Cao R. Spin effects on the EM wave modes in magnetized plasmas. Phys Plasmas 2016;23:112113.

[66] Andreev PA. Separated spin-up and spin-down quantum hydrodynamics of degenerated electrons: Spin-electron acoustic wave appearance. Phys Rev E 2015;91:033111.

[67] Hussain S, Mahmood S. Ion-acoustic shocks in magnetized quantum plasmas with relative density effects of spin-up and spin-down degenerate electrons. Phys Plasmas 2017;24:102.

[68] Kumar P, Singh S, Ahmad N. Conversion efficiency of even harmonics of whistler pulse in quantum magnetoplasma. Laser Part Beams 2019;37:5–11.

[69] Kumar P, Ahmad N. Surface plasma wave in spin-polarized semiconductor quantum plasma. Laser Part Beams 2020;38:159–164.

[70] Kumar P, Singh S, Ahmad N. Beam-plasma streaming instability in spin polarized quantum magnetoplasma. Phys Scr 2020;95:075604.

[71] Walser MW, Keitel CH. Spin-induced force in intense laser-electron interaction. J Phys B At Mol Opt Phys 2000;33:L221.

[72] Qian Z, Vignale G. Spin dynamics from time-dependent spin-density-functional theory. Phys Rev Lett 2002;88:056404.

[73] Liboff RL. Spin and orbital angular momentum. Europhys Lett 2004;68:577.

[74] Marklund M, Brodin G. Dynamics of spin-$\frac{1}{2}$ quantum plasmas. Phys Rev Lett 2007;98:025001.

[75] Brodin G, Marklund M. Spin magnetohydrodynamics. New J Phys 2007;9:277.

[76] Uzma C, Zeba I, Shah HA, Salimullah M. Stimulated Brillouin scattering of laser radiation in a piezoelectric semiconductor: Quantum effect. J Appl Phys 2009;105:013307.

[77] Zeba I, Uzma C, Jamil M, Salimullah M, Shukla PK. Colloidal crystal formation in a semiconductor quantum plasma. Phys Plasmas 2010;17:032105.

[78] Shukla PK, Eliasson B. Novel attractive force between ions in quantum plasmas. Phys Rev Lett 2012;108:165007.

[79] Moldabekov Z, Schoof T, Ludwig P, Bonitz M, Ramazanov T. Statically screened ion potential and Bohm potential in a quantum plasma. Phys Plasmas 2015;22:102106.

[80] Haskins JF, Hickman JS. A derivation and tabulation of the piezoelectric equations of state. J Acoust Soc Am 1950;22:584–588.

[81] Bechmann R. The linear piezoelectric equations of state. Br J Appl Phys 1953;4:210.

[82] Hutson AR, White DL. Elastic wave propagation in piezoelectric semiconductors. J Appl Phys 1962;33:40–47.

[83] Brodin G, Mishra AP, Marklund M. Spin contribution to the ponderomotive force in a plasma. Phys Rev Lett 2010;105:105004.

[84] Brodin G, Marklund M. Spin magnetohydrodynamics. New J Phys 2007;9:277.

[85] Bonitz M, Moldabekov ZA, Ramazanov TS. Quantum hydrodynamics for plasmas—Quo vadis? Phys Plasmas 2019;26:090601.

[86] Bonitz M, Filinov A, Böning J, Dufty JW. Introduction to Complex Plasmas. Springer, Berlin, Heidelberg, 2010.

[87] Ghosh S, Khare P. Effect of density gradient on the acousto-electric wave instability in ion-implanted semiconductor plasmas. Acta Phys Pol A 2006;109:187–197.

[88] Nag BR. Theory of Electrical Transport in Semiconductors. Pergamon Press, Oxford, 1972.

[89] Hassel M, Kwok HS. Picosecond Phenomena III, Springer Series in Chemical Physics. Springer, New York, 1982.

[90] Kallaher RL, Heremans JJ. Spin and phase coherence measured by antilocalization in nnn-InSb thin films. Phys Rev B 2009;79:075322.

[91] Seeger K. Semiconductor Physics. Springer-Verlag, New York, 1973.

[92] Arlt G, Quadflieg P. Piezoelectricity in III-V compounds with a phenomenological analysis of the piezoelectric effect. Phys Status Solidi B 1968;25:323–330.

# Figure Captions

Figure 1. Variation of $\omega/\omega_p$ with $kc/\omega_p$ in piezoelectric semiconductor quantum plasma and in absence of quantum effects $(\hbar = 0)$ for $n_{0e} = 10^{26} m^{-3}$, $\beta = 0.054 C/m^2$.

Figure 2. Variation of $\omega/\omega_p$ with $kc/\omega_p$ for different value of spin polarization $\eta$ for $n_{0e} = 10^{26} m^{-3}$, $\beta = 0.054 C/m^2$.

Figure 3. Variation of $\omega/\omega_p$ with $kc/\omega_p$ for different value of nondimensional quantum parameter $H$ for $n_{0e} = 10^{26} m^{-3}$, $\beta = 0.054 C/m^2$.

Figure 4. Variation of $\Gamma$ with $\omega$ in piezoelectric semiconductor quantum plasma and in absence of quantum effects $(\hbar = 0)$ for $n_{0e} = 10^{26} m^{-3}$, $\beta = 0.054 C/m^2$.

Figure 5. Variation of $\Gamma$ with $\omega$ for different value of spin polarization $\eta$ for $n_{0e} = 10^{26} m^{-3}$, $\beta = 0.054 C/m^2$.

Figure 6. Variation of $\Gamma$ with $\beta$ for different value of nondimensional quantum parameter $H$ for $n_{0e} = 10^{26} m^{-3}$, $\beta = 0.054 C/m^2$.

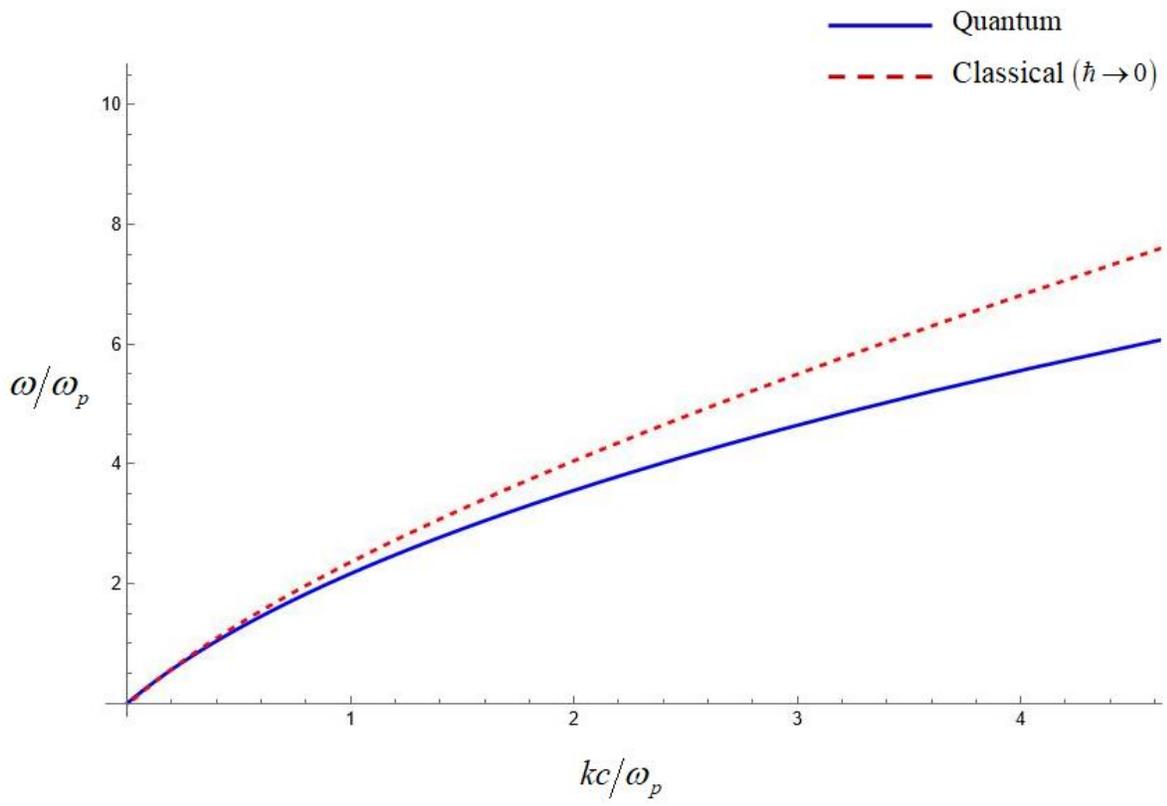

**Fig. 1.**

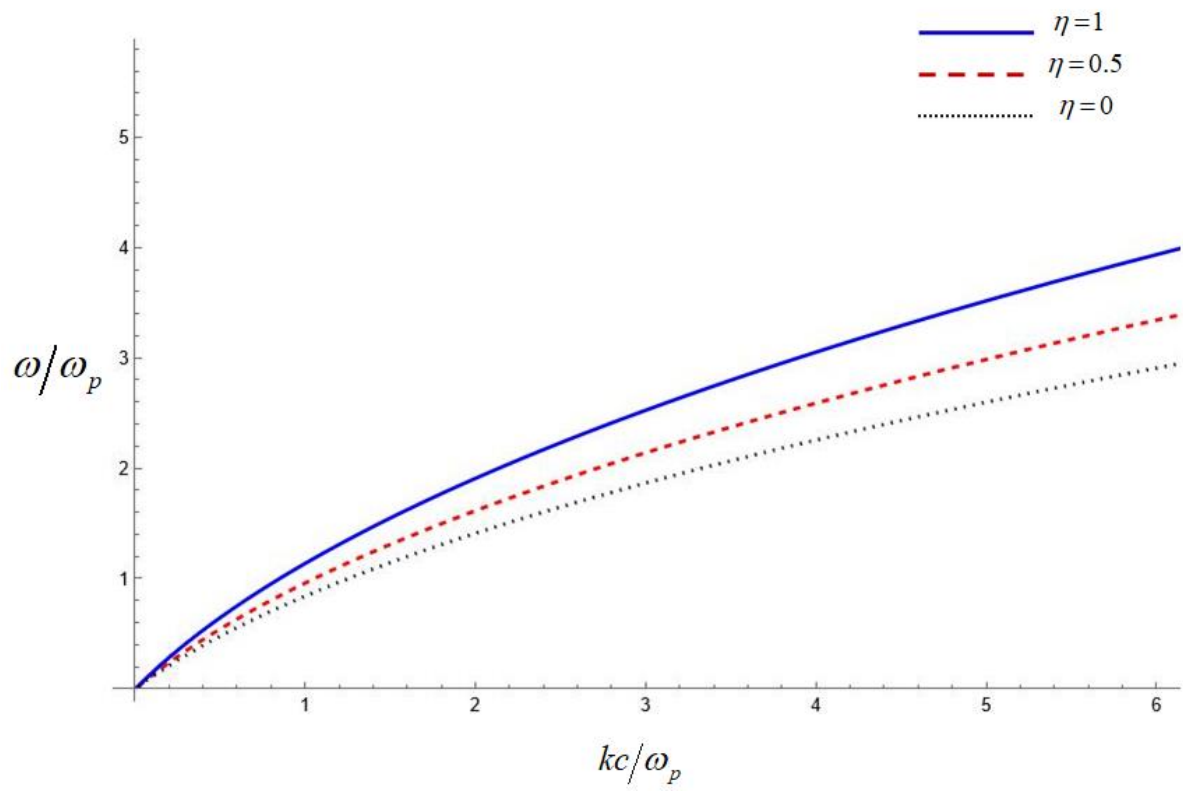

**Fig. 2.**

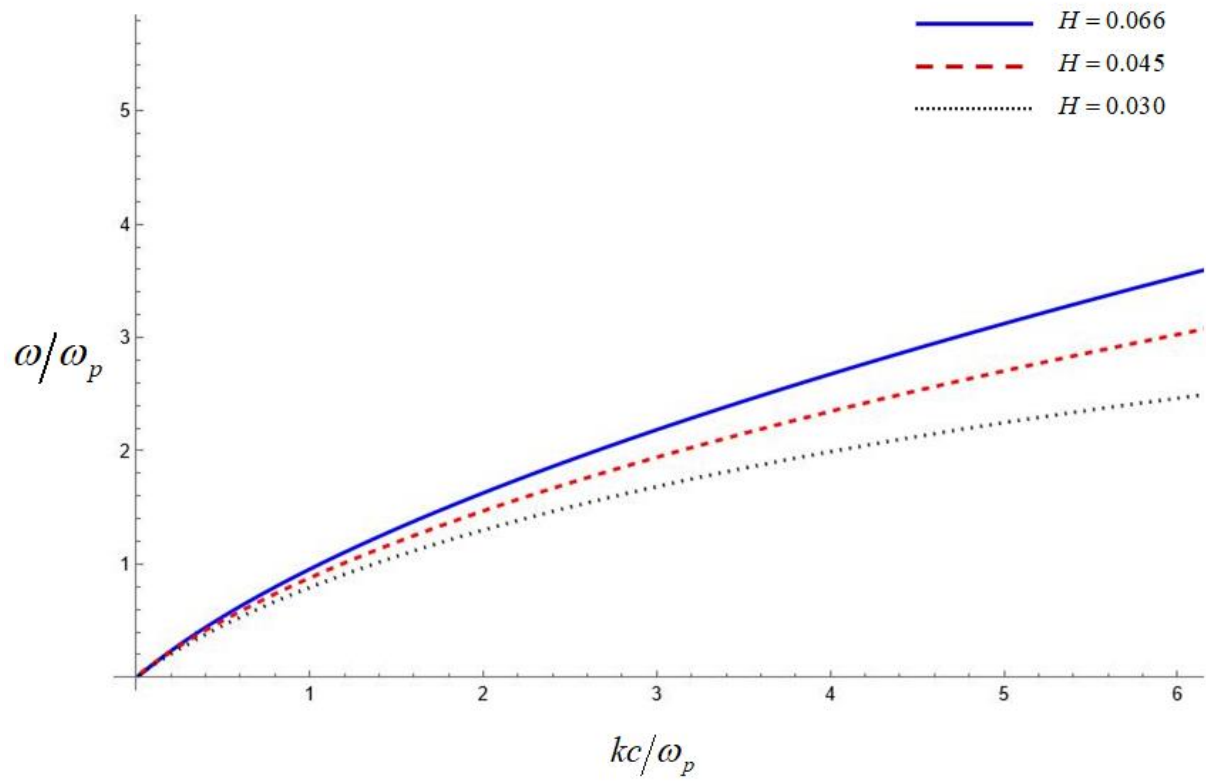

**Fig. 3.**

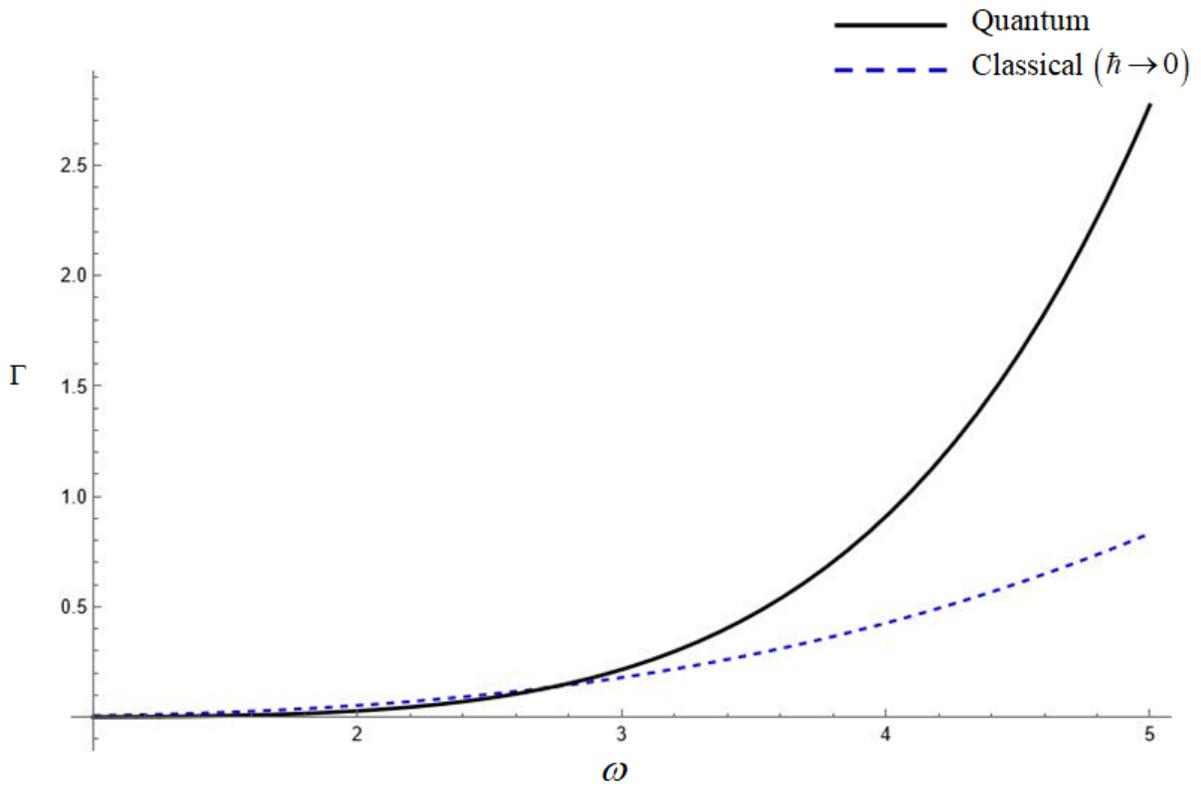

**Fig. 4.**

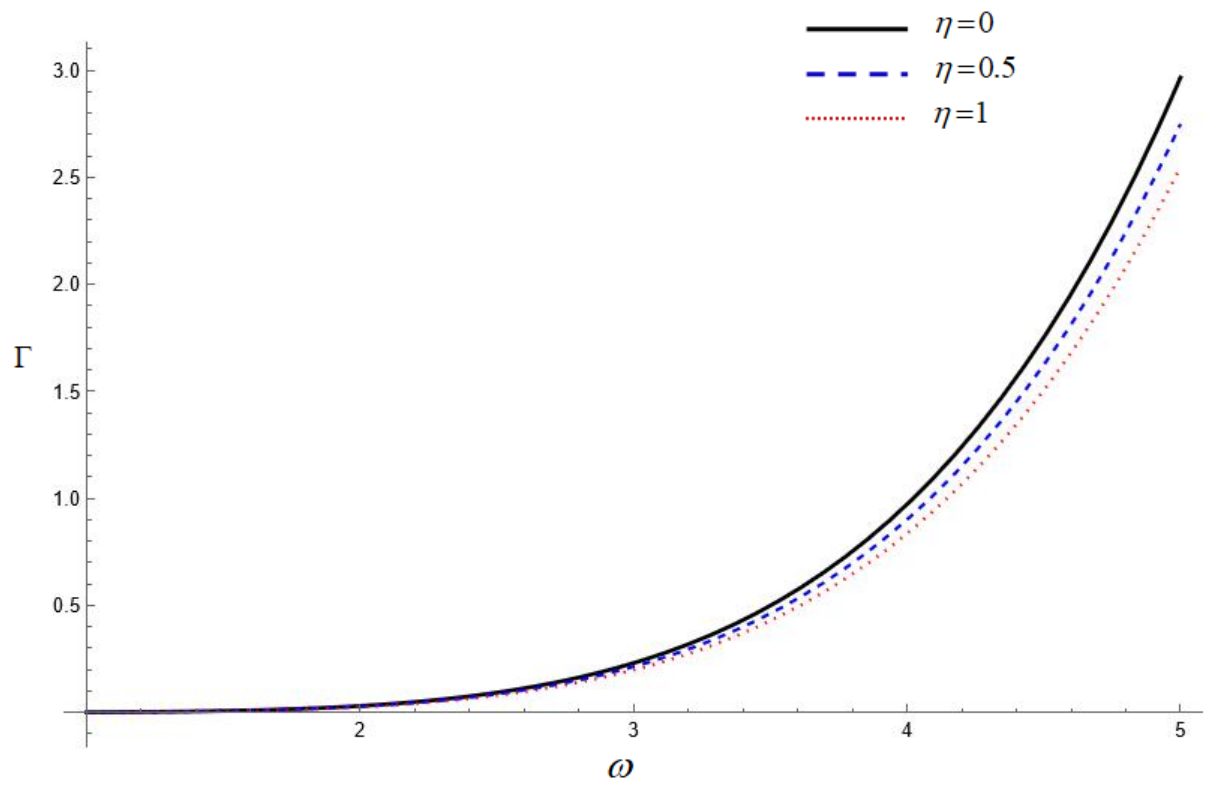

**Fig. 5.**

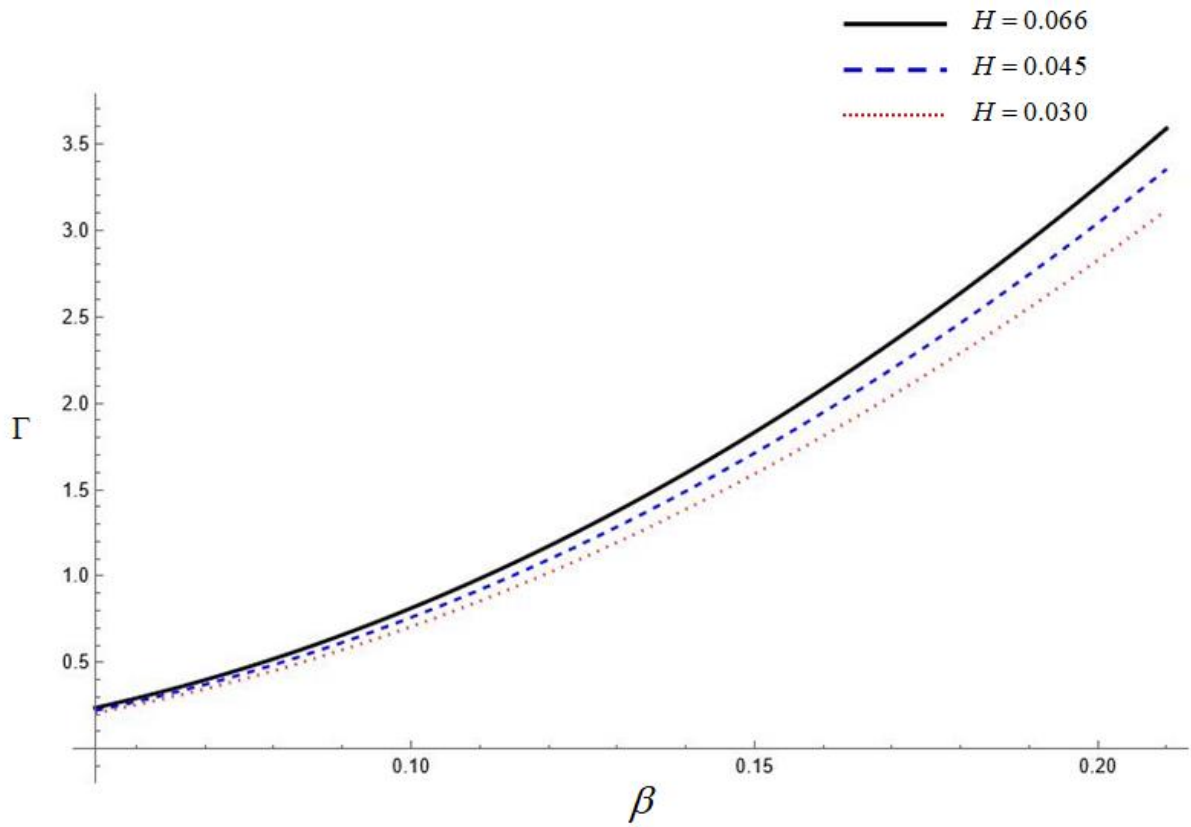

**Fig. 6.**